\RequirePackage{fix-cm}
\RequirePackage{lineno} 
\documentclass[
smallextended,
]{svjour3}       
\smartqed  

\usepackage{graphicx}
\usepackage{dcolumn}
\usepackage{bm}

\newcommand{\hide}[1]{} 

\usepackage{amsmath}            
\usepackage{amssymb}            
\usepackage{latexsym}
\usepackage{subfigure}
\usepackage{rotating}
\usepackage{graphicx}
\usepackage{natbib}

\newcommand{\eg}{e.g., }

\newcommand{\ie}{i.e., }

\newcommand{\sect}[1]{Section \ref{s:#1}}
\newcommand{\sects}[2]{Section \ref{s:#1} and \ref{s:#2}}

\newcommand{\eqn}[1]{Eq.\ (\ref{e:#1})}

\newcommand{\fig}[1]{Fig.\ \ref{fig:#1}}
\newcommand{\figtwo}[2]{Figs.\ \ref{fig:#1} and \ref{fig:#2}}

\newcommand{\Fig}[1]{Figure \ref{fig:#1}}

\begin{document}
\graphicspath{{Figs/}{Figs/PDF}}

\title{Granular Shear Flow in Varying Gravitational Environments}

\author{N. Murdoch         \and
        B. Rozitis \and
        S. F. Green \and
        T-L de Lophem \and
        P. Michel \and
        W. Losert}

\institute{N. Murdoch \at
 Laboratoire Lagrange, UMR 7293, Universit\'{e} de Nice Sophia-Antipolis, CNRS, Observatoire de la C\^{o}te d'Azur, 06300 Nice, France and Planetary and Space Sciences, Department of Physical Sciences, The Open University, Milton Keynes, MK7 6AA\\
\email{murdoch@oca.eu}
\and
B. Rozitis and S. F. Green \at
Planetary and Space Sciences, Department of Physical Sciences, The Open University, Milton Keynes, MK7 6AA
\and
P. Michel \at
 Laboratoire Lagrange, UMR 7293, Universit\'{e} de Nice Sophia-Antipolis,  CNRS, Observatoire de la C\^{o}te d'Azur, 06300 Nice, France
 \and
W. Losert \at
 Institute for Physical Science and Technology, and Department of Physics,  University of Maryland,  College Park, MD 20742, USA \\
 \email{wlosert@umd.edu}
  }

\date{Published online in Granular Matter: 22 February 2013}
                       
                        \maketitle
   
\begin{abstract}

Despite their very low surface gravities, asteroids exhibit a number of different geological processes involving granular matter. Understanding the response of this granular material subject to external forces in microgravity conditions is vital to the design of a successful asteroid sub-surface sampling mechanism, and in the interpretation of the fascinating geology on an asteroid. 

We have designed and flown a Taylor-Couette shear cell to investigate granular flow due to rotational shear forces under the conditions of parabolic flight microgravity.  The experiments occur under weak compression.  First, we present the technical details of the experimental design with particular emphasis on how the equipment has been specifically designed for the parabolic flight environment. Then, we investigate how a steady state granular flow induced by rotational shear forces differs in varying gravitational environments. We find that the effect of constant shearing on the granular material, in a direction perpendicular to the effective acceleration, does not seem to be strongly influenced by gravity. This means that shear bands can form in the presence of a weak gravitational field just as on Earth. 

\keywords{Microgravity \and Shear \and Experiment \and Granular \and Asteroid}

\end{abstract}


\section{Introduction}\label{s:intro}

Asteroids are covered by granular material that can range in size from a few microns (dust) or few hundreds of microns (sand) to a few centimeters or meters (gravels, pebbles, boulders). This superficial layer extends to variable depths and essentially results from impact processes via excavation, fragmentation and ejection of material. Granular materials (regolith) on the surfaces of planets and moons in the Solar System may be also produced via other geological processes such as volcanic activity, erosion and transport, and are extremely common at the surface of all solid bodies. Therefore, the dynamics of granular materials are involved in planetary evolution but appear to also be critical for the design and/or operations of landers, sampling devices and rovers.

Our experiment was designed to study the dynamics of a granular material subject to shear forces in a microgravity environment using a Taylor-Couette shear cell. Couette flow, a term named after the French physicist Maurice Couette, was first used in the field of fluid dynamics \citep{couette1888}. It refers to the laminar flow (a streamlined flow in parallel layers within which there are no cross currents or eddies) of a fluid between two parallel plates, one of which is moving relative to  the other.  In this simple Couette geometry the two plates cannot extend infinitely in the flow direction, and so, in order to study shear-driven flows, Sir Geoffrey Taylor created a Couette shear cell using rotating co-axial cylinders \citep{taylor23}. The circular Couette shear cell, also known as the Taylor-Couette shear cell, has been used in countless experiments in fluid dynamics and, more recently, in studies of granular material on the ground.

Although a similar experimental set-up can be used to study both fluids and granular materials the two media react very differently to shear stresses. Neglecting possible edge effects and effects due to the curvature of the shear cell, the Couette (angular velocity) profile of a Newtonian fluid (a fluid for which there is a linear relationship between the stress and the strain) is a decreasing function across the entire shear cell width. However, in a similar experiment, a granular material develops a shear band;  a narrow zone of large relative particle motion bounded with essentially rigid regions. Almost all of the energy input into the granular system by the inner cylinder is dissipated by friction within this narrow region producing large velocity gradients. Shear bands mark areas of flow, material failure, and energy dissipation, and as such they are important in many geophysical processes \citep[\eg][]{chemenda12, jiang95, gapais89}.

Various different configurations of the Taylor-Couette shear cell have been used to study the effect of shear stresses on granular materials experimentally. For example, \cite{bocquet01} shear a granular material in a Taylor-Couette cell with a rotating inner cylinder and stationary outer cylinder. The inner cylinder is connected to the motor via a spring that allows either stick-slip motion or steady shearing depending on the parameters. By performing force measurements they determined that the shear force acting on the moving cylinder is independent of shearing velocity. The dynamics of individual particles were also investigated by measuring the mean velocity of particles on the top-surface. The authors concluded that the normalised velocity profiles are independent of shearing velocity and of the type of motion of the inner cylinder (\ie stick-slip or continuous sliding).  \cite{tardos98} shear a fine, dry powder in a Couette cell. They measured the torque generated on the rough, inner cylinder and found that, in a column of granular material undergoing continuous shearing, normal and shear stresses increase linearly with depth.  \cite{khosropour97} investigate the size segregation of a binary mixture of spherical glass particles in a Taylor-Couette geometry, where the cylinders are made of smooth glass and the flow is generated by the shearing motion of the inner cylinder. The trajectories of 1, 2, and 3 mm glass particles, placed at the bottom of the cell, were followed as they moved through a 1 mm medium. The authors observed that the larger particles rose to the top and remained on the surface. 

Nothing is known regarding the same process in the absence of gravity or in the presence of a weak gravity field. To reduce the ambient gravitational acceleration during our experiments we use the microgravity environment available on board a parabolic flight (Novespace A300 Zero-G aircraft). During each parabola of a parabolic flight there are three distinct phases: a $\sim$20 second $\sim$1.8 $g$ ($g$ being the gravitational acceleration at the surface of the Earth) injection phase as the plane accelerates upwards, a $\sim$22 second microgravity phase as the plane flies on a parabolic trajectory (during this period the pilot carefully adjusts the thrust of the aircraft to compensate for the air drag so that there is no lift), and finally, a $\sim$20 second $\sim$1.8 $g$ recovery phase as the plane pulls out of the parabola. Between the end of one parabola and the start of the next there is a 2 minute 1 $g$ rest period. This means the starts of the parabolas are 3 minutes apart.  After each set of 5 parabolas it is standard procedure to take a longer 1 $g$ rest of 4-8 minutes. Our experiments were performed with the company Novespace (in France). During one flight there are 31 parabolas and each flight campaign normally consists of 3 flights. This means that there are 93 parabolas in one flight campaign giving approximately 30 minutes of (disconnected) microgravity in total.

We perform constant shear rate experiments with a Taylor-Couette shear cell to investigate how a steady state granular flow, induced by rotational shear forces, differs in varying gravitational environments. The experiments, which occur under weak compression, also allow us to investigate if the formation of shear bands differs between the two gravitational regimes. All of these experiments were performed in the framework of the competitive ESA program `Fly your Thesis'.

In \sect{hardware} we present the technical details of the hardware developed for our experiment with particular emphasis on how the equipment has been designed specifically for the parabolic flight environment. In \sects{procedures}{pp} we present the experimental procedures and we report on some tests that were performed prior to the microgravity flight campaign. Then, finally, in \sect{results} we present the first results examining the constant shear rate experiments performed with our shear cell on the ground and during the parabolic flights.

\section{Experimental design}\label{s:hardware}

The AstEx (ASTeroid EXperiment) experiment flew in the European Space Agency (ESA) 51st Microgravity Research Campaign in November 2009 as part of ESAÕs `Fly your Thesis' programme \citep{callens11}.  Our experiment uses a Taylor-Couette shear cell (modified for the parabolic flight environment) that is housed, along with the mechanical and electrical components, inside an experimental rack. This section presents the technical details of the hardware developed for the AstEx experiment with particular emphasis on how the equipment has been designed specifically for the parabolic flight environment. 

\subsection{The AstEx shear cell}\label{s:MM_cell}

\begin{figure*} 
\centering
\subfigure[]{
 \includegraphics[width=0.6\columnwidth]{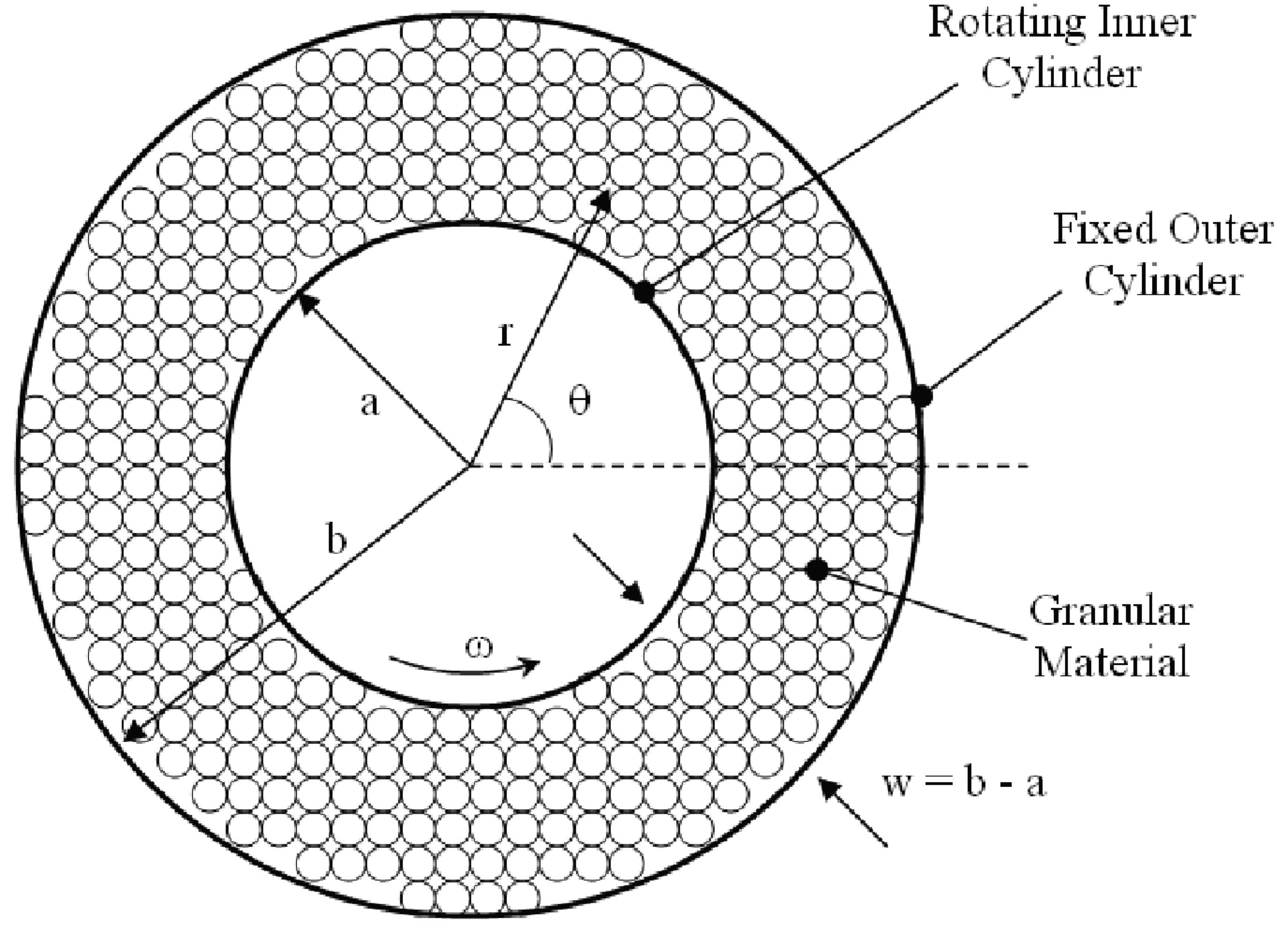}
}
 \caption{The Taylor-Couette Geometry. $a$ = Inner Cylinder Radius, $b$ = Outer Cylinder Radius, $w$ = Width of Shear Region, $r$ = Radial Distance, $\theta$ = Angular Distance, $\omega$ = Inner Cylinder Angular Velocity.  Image adapted from \cite{toiya06}.}
 \label{fig:Taylor_Couette}
\end{figure*}

\begin{figure*} 
\centering
\subfigure[]{
 \includegraphics[width=0.6\columnwidth]{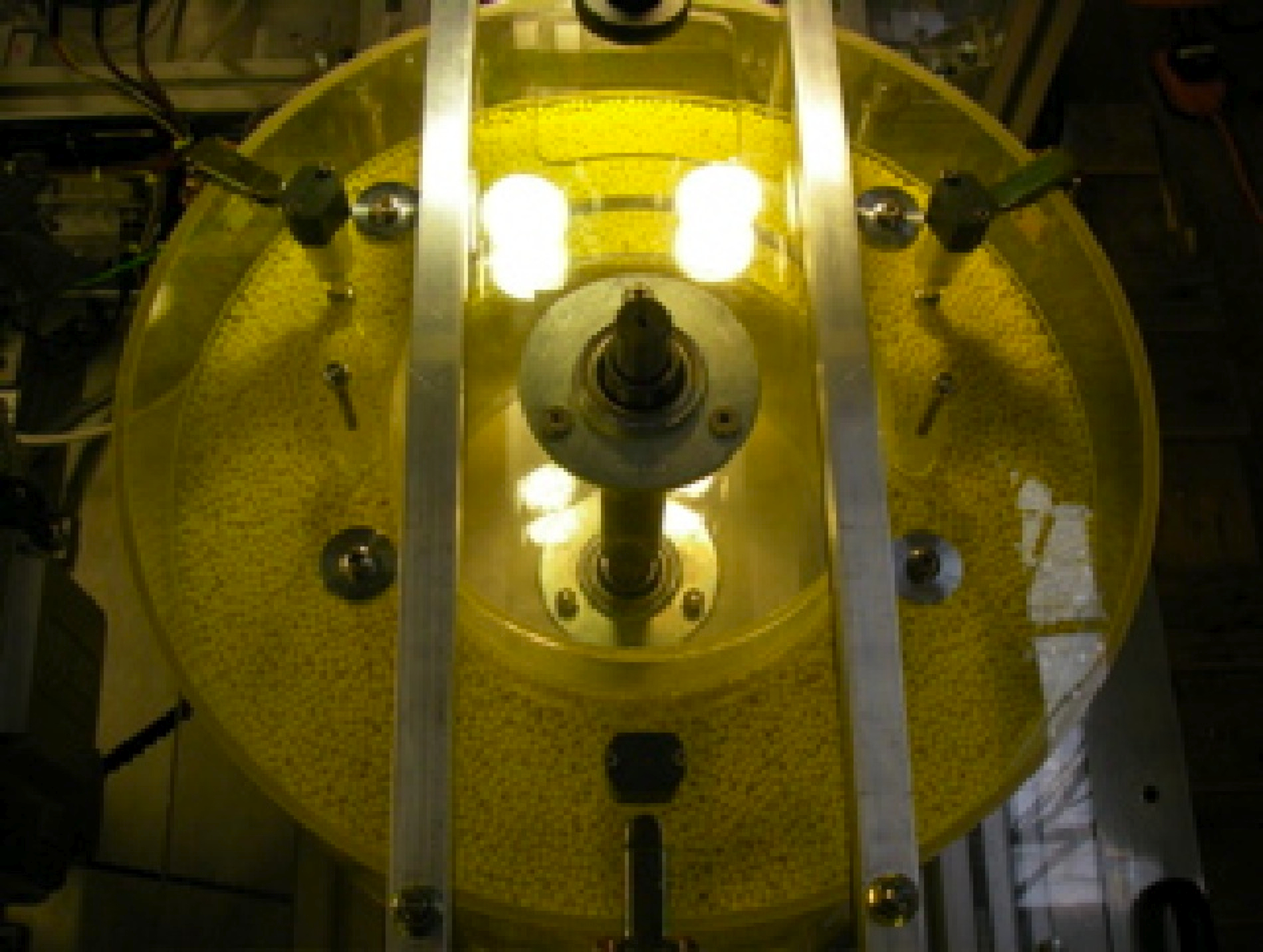}
}
 \caption{A photograph of the AstEx Taylor-Couette shear cell.}
 \label{fig:astex_cell}
\end{figure*}

Our experiment uses the simplest Taylor-Couette geometry as shown in \fig{Taylor_Couette}.  There are two concentric cylinders made from cast Acrylic tubes; the outer cylinder has an inner radius of 195 mm and the inner cylinder has an outer radius of 100 mm.  Both the inner and outer cylinders are 200 mm in height. The outer cylinder is fixed and its inner surface is rough with a layer of particles, the outer surface of the inner cylinder is also rough but it is free to rotate, and the floor between the two cylinders is smooth and fixed in place. The gap between the two cylinders is filled to a height of 100 mm with spherical soda lime glass beads (grain diameter, d $=$ 3 or 4 mm; density, $\rho$ $=$  2.55 g cm$^{-3}$) upon which the rotating inner cylinder applies shear stresses.  \Fig{astex_cell} is a photograph of a single shear cell.

\begin{center}
  [FIGURE \ref{fig:Taylor_Couette} GOES HERE]
\end{center}

\begin{center}
  [FIGURE \ref{fig:astex_cell} GOES HERE]
\end{center}

\begin{figure*} 
\centering
\subfigure[]{
 \includegraphics[width=0.6\columnwidth]{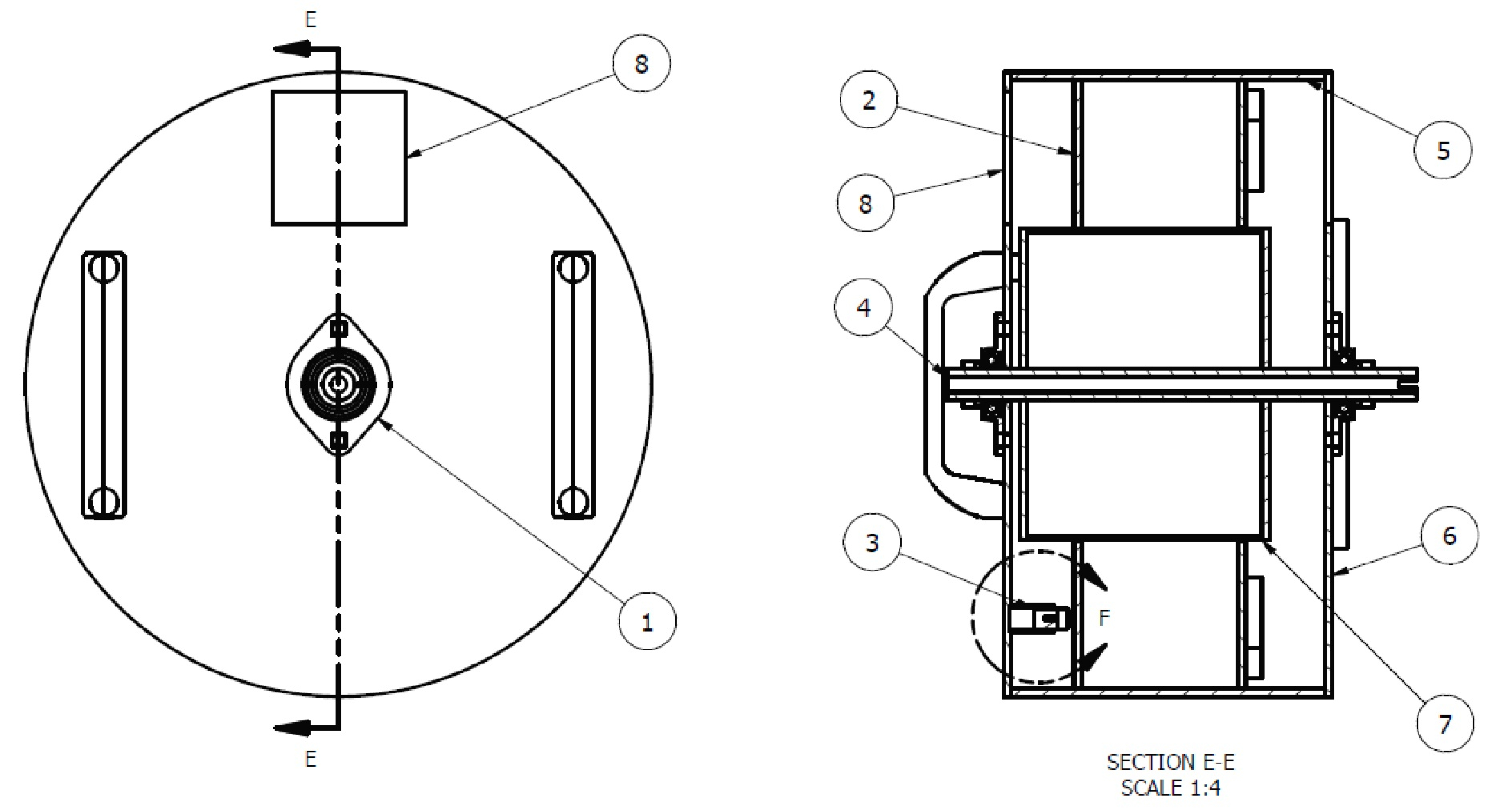}
}
 \caption{Internal workings of the shear cell. (1) Housing unit with bearing (2) Movable pressure disk (3) Pressure springs (4) Steel shaft  (5) Fixed outer cylinder (6) Fixed bottom plate (7) Rotating inner cylinder  (8) Camera viewport. }
 \label{fig:internal_workings}
\end{figure*}

\begin{figure*} 
\centering
\subfigure[]{
 \includegraphics[width=0.6\columnwidth]{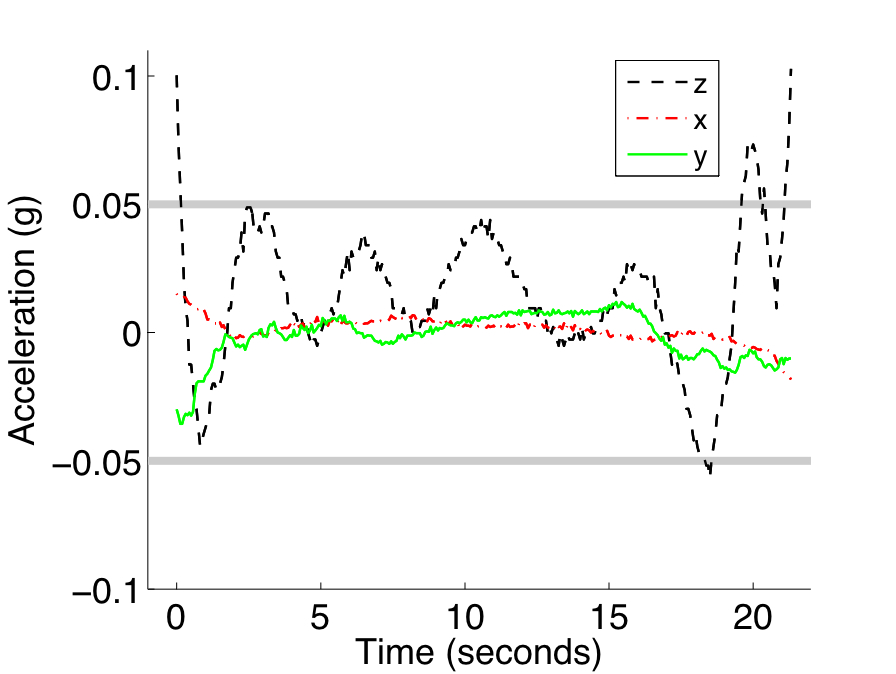}
}
 \caption{Example data showing the acceleration in the z-, x- and y-directions during the microgravity part of the parabola \ie while the acceleration in the z-direction is $<$ 0.1 $g$. Novespace aim to maintain the fluctuations to within the limits of  0 $\pm$ 0.05 $g$, indicated by the horizontal grey lines. These data were provided by Novespace during the flight campaign.}
 \label{fig:gravity}
\end{figure*}

We confine the granular material by exerting a very low positive force on the top surface of the beads using a \emph{pressure plate}: a sprung loaded movable, transparent disk (\fig{internal_workings}).  This ensures all sidewalls can sustain forces to contain the particles.  The microgravity environment on a parabolic flight is not perfect; there are small fluctuations that Novespace aim to maintain within the limits of  0 $\pm$ 0.05 $g$ (\fig{gravity}). This means that, at times, the experiment will experience small amounts of negative $g$, which will lead to dilation of the granular material. The mass of beads contained in the shear cell, assuming an initial packing fraction ($\psi$) of 0.6 and a filling height of 100 mm is 13.47 kg. As the magnitude of the maximum negative acceleration possible is 0.05 $g$, the force required to compensate for this acceleration, and prevent dilation of the granular material, is 6.6 N. This force is distributed between 3 springs; each spring provides a force of 2.2 N at the normal filling height ($\psi$ = 0.6), a force of 0 N at the minimum filling height ($\psi$ = 0.645) and a maximum force of 4.4 N at the maximum filling height \citep[$\psi$ = 0.555; ][]{onoda90}. This is equivalent to pressure variations of between 0 and 149 Pa, assuming that the pressure is equally distributed over the entire area of the plate.

\begin{center}
  [FIGURE \ref{fig:internal_workings} GOES HERE]
\end{center}

\begin{center}
  [FIGURE \ref{fig:gravity} GOES HERE]
\end{center}

\subsection{The AstEx experimental rack}

\begin{figure*} 
\centering
\subfigure[]{
\includegraphics[width=0.45\columnwidth]{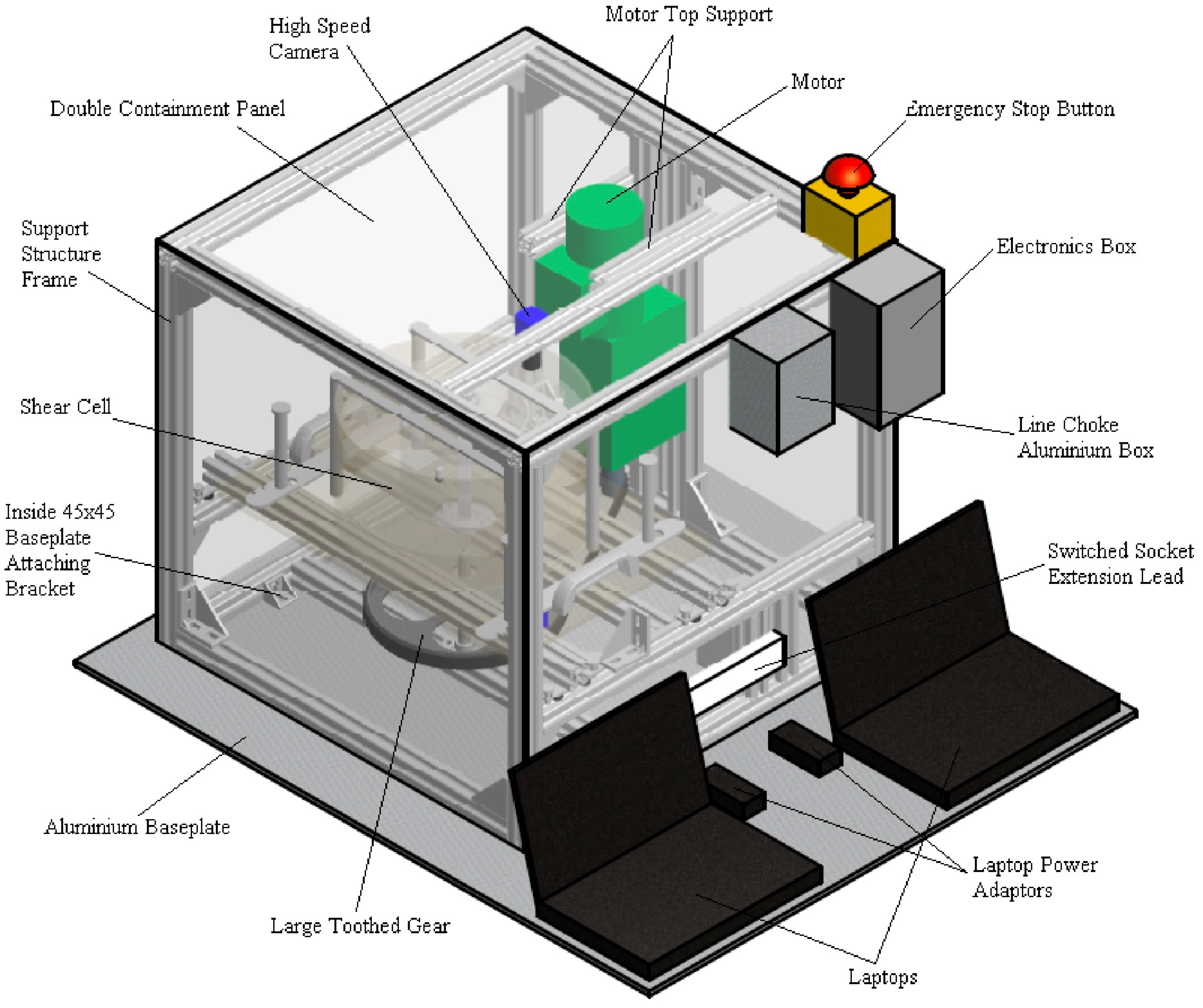}
}
\subfigure[]{
\includegraphics[width=0.45\columnwidth]{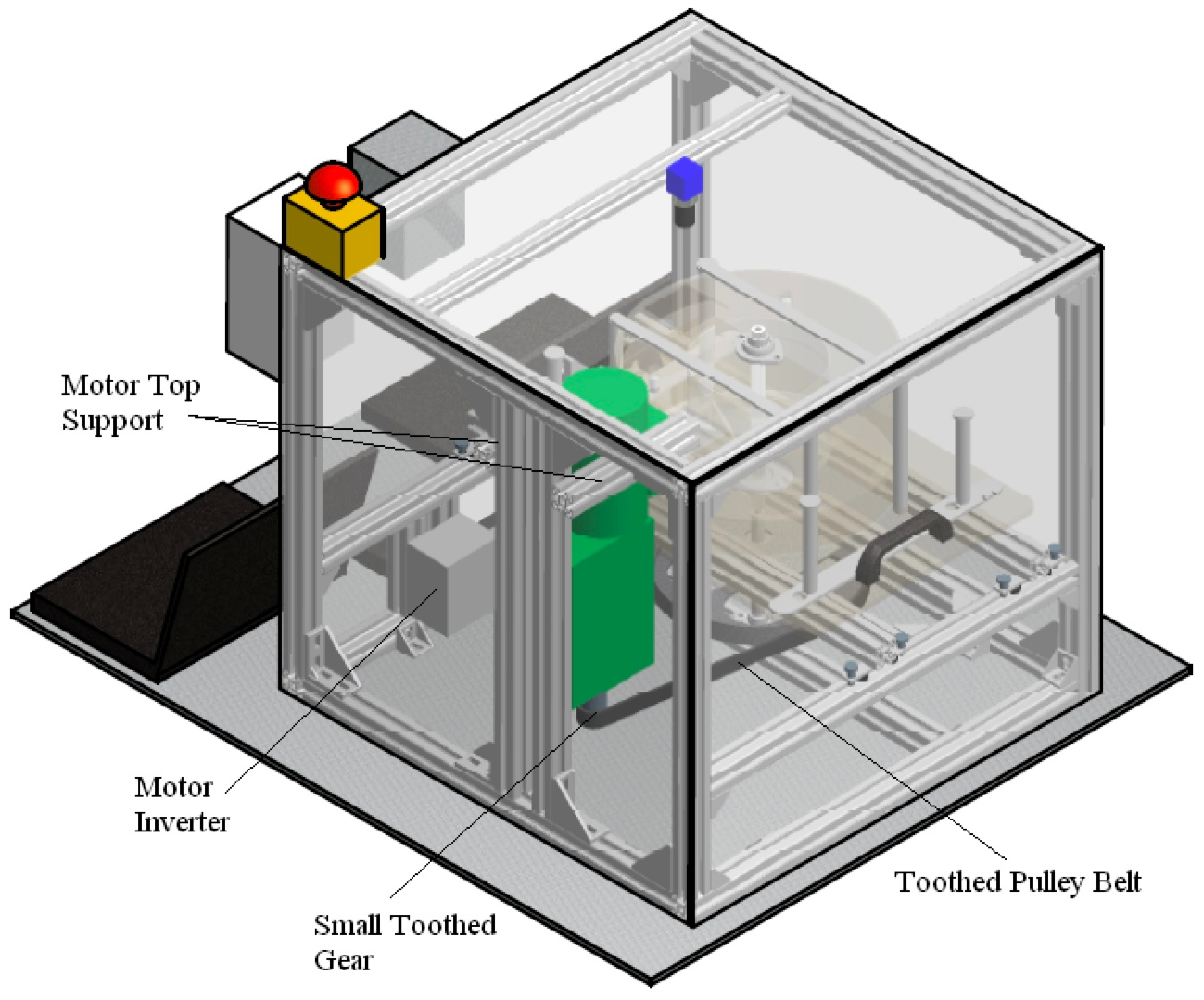}
}
 \caption{Design drawings of the AstEx experimental rack. (a) Front view; (b) Rear view.}
   \label{fig:rack}
\end{figure*}

\Fig{rack} shows how our Taylor-Couette shear cell is mounted inside the A300 Zero-G aircraft for testing during the parabolic flights. The experiment rack consists of two parts: a test compartment, and a laptop work station. The test compartment is where the experiments take place and it contains one shear cell.  Situated next to the test compartment is the laptop work station where two laptops are mounted to allow two experimenters to control the various components of the experimental hardware, to perform the experiments and to collect and store data. The laptops were specially requested without freefall sensors (a standard feature in most laptop hard drives). The entire experiment rack is 1006 $\times$ 1250 $\times$ 750 mm in size, and has a total mass of $\sim$170 kg.

\begin{center}
  [FIGURE \ref{fig:rack} GOES HERE]
\end{center}

\begin{figure*} 
\centering
\subfigure[]{
 \includegraphics[width=0.6\columnwidth]{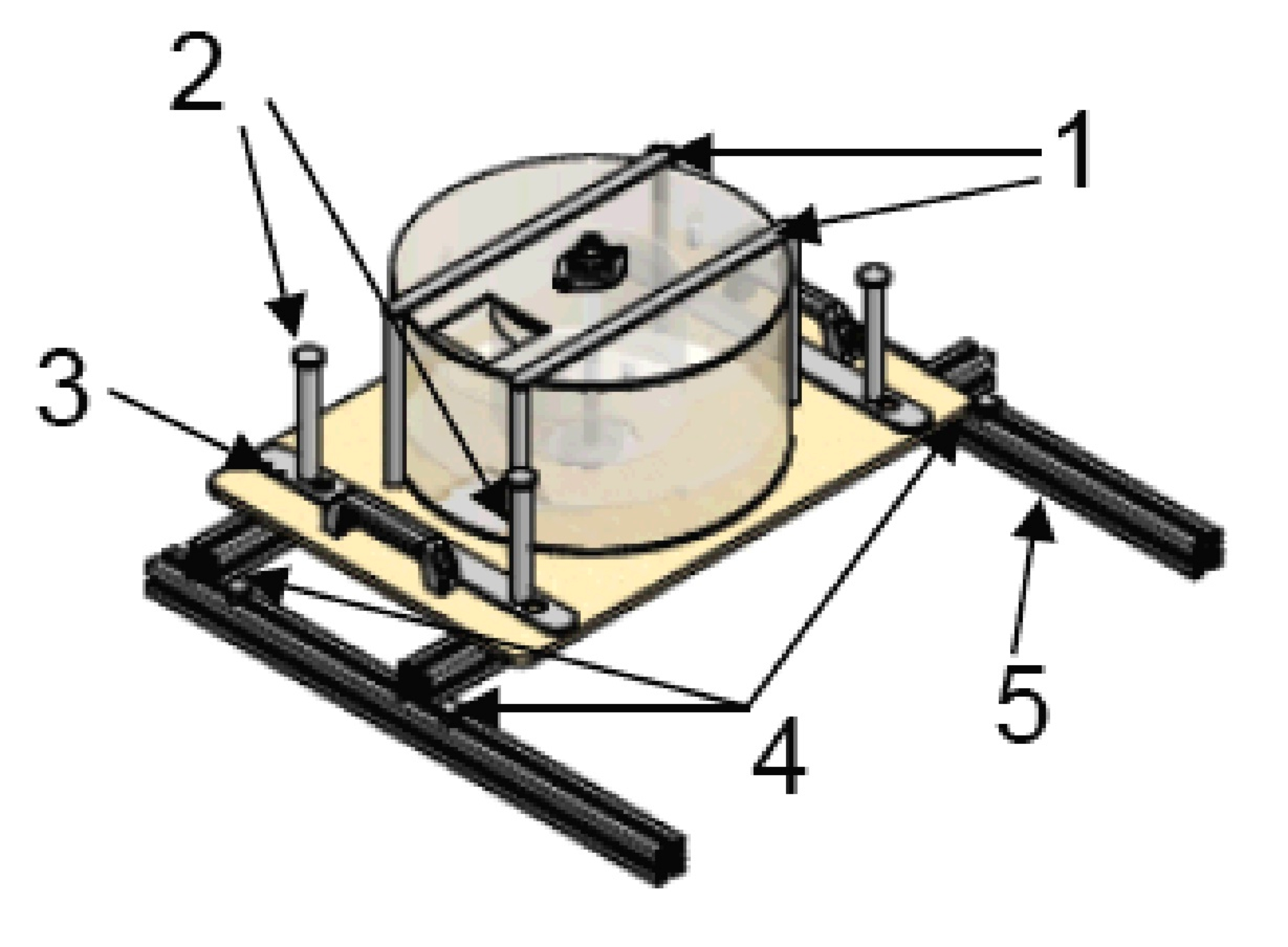}
}
 \caption{Shear cell mounting. (1) Shear cell securing bars (2) Guide rods (3) Sliding locking bars (4) Silent blocks (5) Support structure.}
 \label{fig:mounting}
\end{figure*}

The shear cells were designed so that they could be easily exchanged between flights when the plane is on the ground. This allows different shear cells to be flown during the parabolic flight campaign. The shear cells are mounted on a polycarbonate base before being placed into position in the experiment rack (\fig{mounting}). Only one shear cell can be attached to the experiment rack at one time. The polycarbonate base features two handles used to shake the shear cell between experiments.  The inner cylinder has a steel shaft running through the centre and is attached to the outer cylinder by two bearings contained inside a housing unit. This steel shaft attaches to the driving shaft and transmits the torque produced by the motor to the inner cylinder. A Watt Drive HU50C 64K4 inline helical geared three phase motor is used, controlled by a Moeller DF51-322-025 three phase inverter. 

To minimise the effects of aircraft vibrations we attempt to isolate the shear cell from the aircraft. This is done by mounting silent blocks (a type of vibration isolator made of rubber) between the two strut profiles on which the shear cell is resting and the rest of the support structure frame (\fig{mounting}).

\begin{center}
  [FIGURE \ref{fig:mounting} GOES HERE]
\end{center}

\subsection{Data collection}\label{s:data_collection}

Two high-speed cameras (Matrix Vision Blue Fox 120aG) image the top and bottom layers of glass beads in the shear cell at $\sim$60 frames sec$^{-1}$ (fps) so that the particles do not move more than 1/10 $d$ between consecutive frames (where $d$ is the particle diameter). The cameras, which each have a resolution of 640 $\times$ 480 pixels, are mounted to the experiment rack in the test compartment and image the glass beads through camera viewports built into the shear cells (see \figtwo{internal_workings}{rack}). Four energy-saving reflector lamps are mounted next to the cameras (two for each camera) to illuminate the glass beads.   Several thousand images were taken with each camera during each experimental run and saved to the laptop hard-disks within the time between parabolas (\ie in less than two minutes) using our own customised data acquisition software.

\section{Experimental Procedures}\label{s:procedures}

During each parabola the same sequence of events is followed. Approximately 5 seconds before the start of the injection phase of the parabola the high-speed cameras start recording. As soon as the microgravity phase starts the motor is started. The motor continues to run during the $\sim$1.8 $g$ recovery phase. Finally, when the 1 $g$ rest phase starts the motor and high-speed cameras are stopped.  During the 2 minute rest period the shear cell is shaken by hand to create reasonably consistent initial arrangement of glass beads and to attempt to remove any contact networks and possible memory effects from the granular material. This procedure is repeated for each parabola; however, during the longer 4-8 minute rests the motor rotation speed was also adjusted. During the parabolic flight campaign, experiments were performed with the inner cylinder rotating at 0.025, 0.05 and 0.1 rad sec$^{-1}$. 

After the flights, particle tracking was performed using an adaptation of a subpixel-accuracy particle detection and tracking algorithm \citep{crocker96}, which locates particles with an accuracy of approximately $1/10$ pixel.  The raw tracking data was smoothed over time using a local regression weighted linear least squares fit (loess model) and then, using the particle positions in each frame, the average angular velocities of every particle can be computed.  The pixel scale for the images of the top surface was determined using beads that are glued to the top surface of the confining pressure plate (these can be seen in \fig{StackedImages} below). The real separations were measured, centre to centre, with vernier callipers. These beads were then identified in the images and their centres were located allowing the pixel scale to be calculated. For the bottom surface the pixel scale was calculated by choosing several particles in a line from within a crystallised region.  The distance between the centres of all of these particles was combined with the known particle radius to calculate the pixel scale. In the few laboratory-based runs that were also performed with no pressure plate (see \sect{pp}) the pixel scale on the top surface was calculated using bead diameters.

In order to compare experiments at different inner cylinder rotational velocities and with different bead sizes, the velocities are normalised. The mean angular velocity of several experiments of the same type, $\overline{V_{\theta}}$, is plotted in its normalised form, 

\begin{equation}
\overline{V^*_{\theta}} =  \frac{\overline{V_{\theta}}}{\omega}
\label{e:VstarTheta}
\end{equation}

\noindent where $\omega$ is the inner cylinder angular velocity.

\section{Experimental Tests}\label{s:pp}

\begin{figure*} 
\centering
\subfigure[]{
\includegraphics[width=0.45\columnwidth]{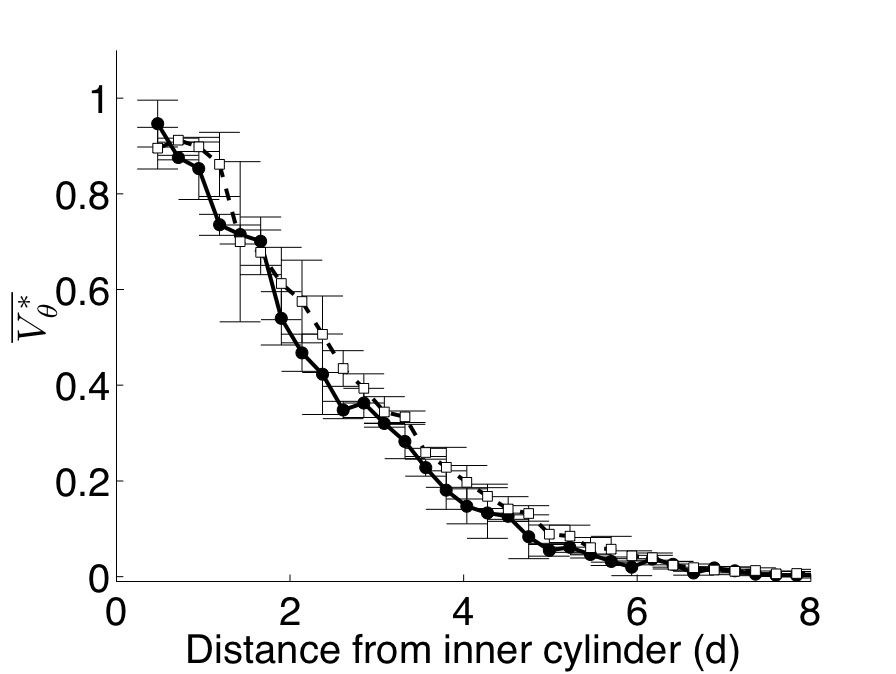}
  \label{fig:PPInfluence4mm4mHz}
}
\subfigure[]{
\includegraphics[width=0.45\columnwidth]{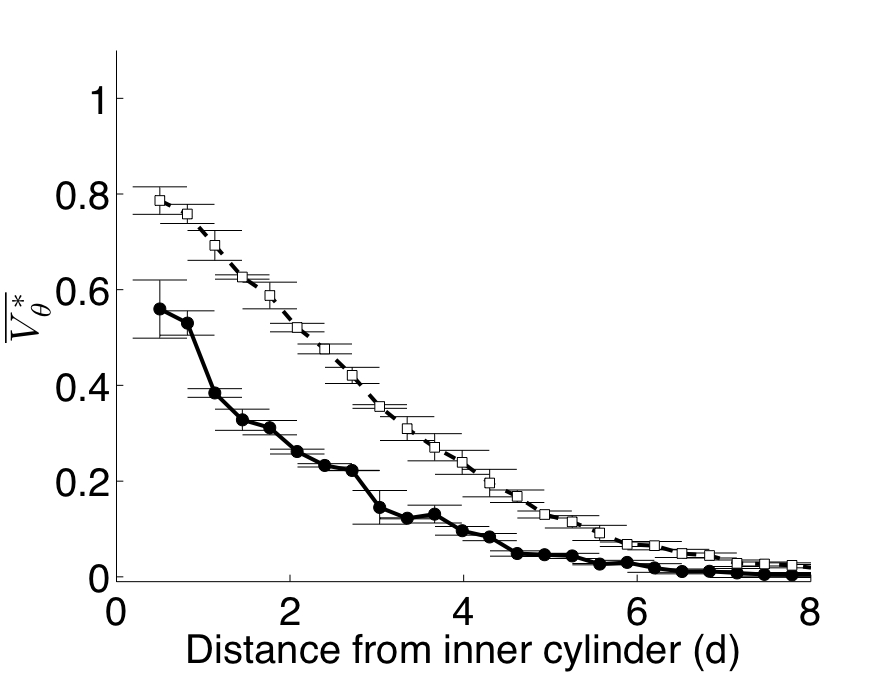}
  \label{fig:PPInfluence3mm4mHz}
}
 \caption{$\overline{V^*_\theta}$ as a function of distance from the inner cylinder for the beads on the top surface of an experiment with (a) 4 mm  and (b) 3 mm beads at an inner cylinder angular velocity of 0.025 rad s$^{-1}$.  Results obtained with the pressure plate (solid line, solid circles) and without the pressure plate (dashed line, open squares) are shown. The error bars represent the standard deviation of $\overline{V^*_\theta}$. The velocity profiles shown only extend up to 8 particle diameters from the shearing surface as further away there is very little particle motion. }
   \label{fig:PPInfluence}
\end{figure*}

As described above, the AstEx experiment had to be modified for the parabolic flight environment. As a result of the modifications to the hardware and experimental set-up, there are a few factors that may influence the measured particle dynamics, namely the presence of the pressure plate. Experiments were performed on the ground with, and without, the pressure plate to determine if the pressure plate had any influence on the particle velocities. \Fig{PPInfluence4mm4mHz} shows the normalised mean angular velocity profiles (\ie $\overline{V^*_{\theta}(r)}$) of the 4 mm particles on the top surface of an experiment with the pressure plate and without the pressure plate with an inner cylinder angular velocity of 0.025 rad s$^{-1}$. The velocity profiles with and without the pressure plate are very similar, although, there are some irregularities in the velocity profiles with the pressure plate. These irregularities seem to occur at similar locations in all experiments and, therefore, may be caused by an inhomegenity in the pressure plate. Part of the pressure plate possibly exerts more pressure on the top layer of particles, or part of the pressure plate is perhaps rougher than the rest. The irregularities also become very slightly more pronounced as the angular velocity increases, which means that the pressure plate has a larger effect on the particles at larger rotation rates.

\Fig{PPInfluence3mm4mHz} shows the normalised mean angular velocity profiles of the 3 mm particles on the top surface of an experiment with the pressure plate and without the pressure plate again with an inner cylinder angular velocity of 0.025 rad s$^{-1}$. Contrary to the 4 mm particle results, the pressure plate appears to have a very strong influence on the mean angular velocity profiles of the 3 mm particles. This may be because the filling heights are slightly different in the two shear cells; the mass of beads to be added to the shear cells was calculated assuming a random packing fraction of $\sim$0.6 and a desired filling height of 100 mm. It is possible that the 3 mm particles have a lower packing fraction than the 4 mm beads and the filling height is, therefore, slightly higher. This would mean that the pressure plate is exerting a slightly larger pressure on the 3 mm beads.  

However, another possible explanation is that the rate at which the particles are decelerated by the pressure plate depends on their radius. Assuming we have one layer of mono-disperse particles (particle radius, $r$, and density, $\rho$), covering a given surface area, $A_s$, and with a 2d packing fraction, $\phi_{2d}$, then the total number of particles in the layer, $N_p$, is given by,

\begin{equation}
N_p = \frac{A_s  \phi_{2d}}{\pi r^2}.
\label{e:NpIn2dLayer}
\end{equation}

\noindent If the total force exerted on the particles by the pressure plate is, $F_N$, then the normal force experienced by each particle ($F_{Np}$) on the surface layer is,

\begin{equation}
F_{Np} = \frac{F_N}{N_p} =  \frac{\pi F_N r^2 }{A_s  \phi_{2d}}.
\label{e:NpIn2dLayer}
\end{equation}

\noindent The frictional force experienced by each particle ($F_{Sp}$) is related to the normal force experienced by each particle via the following relation:

\begin{equation}
 F_{Sp} = \mu_s F_{Np}
 \end{equation}
 
\noindent  where $\mu_s$ is the coefficient of static friction. The resulting deceleration ($a_p$) of each particle in the surface layer due to the pressure plate is given by:

\begin{equation}
a_p =  \frac{F_{Sp}}{M_p} = \frac{3 \mu_s F_N}{4 \phi_{2d} \rho r}
\label{e:Decel}
\end{equation}

\noindent  where the mass of each particle ($M_p$) is,

\begin{equation}
M_p = \frac{4 \pi r^3 \rho}{3}.
\label{e:Mp}
\end{equation}

\noindent Therefore, with constant total normal force, particle density, surface area and packing fraction, the deceleration of each particle should vary inversely with $r$. This means that the smaller (3 mm) particles should be decelerated by the pressure plate more than larger (4 mm) particles. Our experimental results do indeed show that the 3 mm particles 
have a lower velocity in the presence of the pressure plate, whereas the 4 mm particles do not  (\fig{PPInfluence}). However, from these two data sets it is not possible to conclude with certainty that this mechanism is responsible.

\begin{center}
  [FIGURE \ref{fig:PPInfluence} GOES HERE]
\end{center}

\section{Steady state granular flow in varying gravitational environments} \label{s:results}

\begin{figure*} 
\centering
\subfigure[]{
\includegraphics[width=0.45\columnwidth]{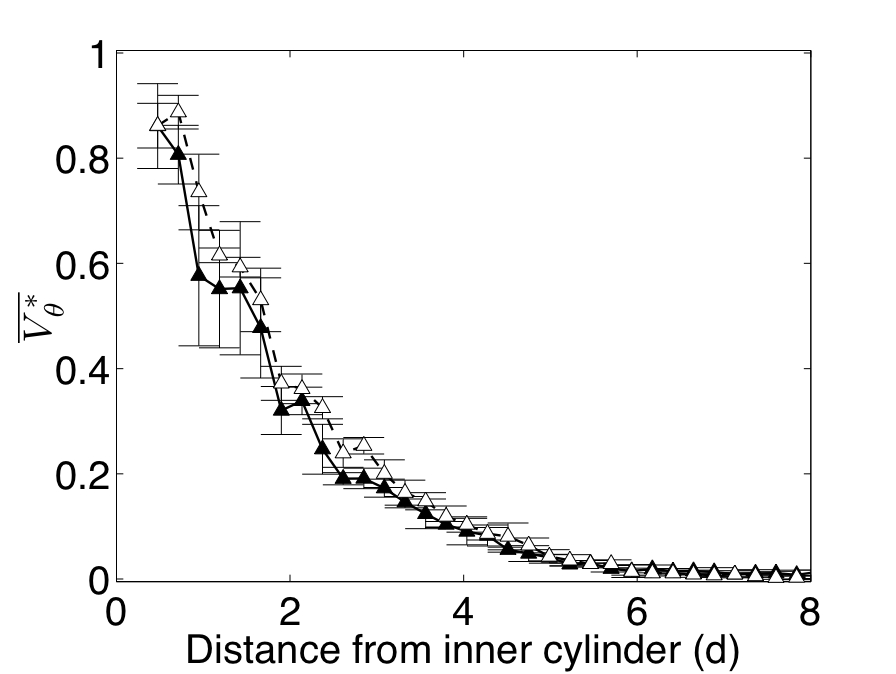}
  \label{fig:Cam2_4mm_MeanAngVelProfs1g}
}
\subfigure[]{
\includegraphics[width=0.45\columnwidth]{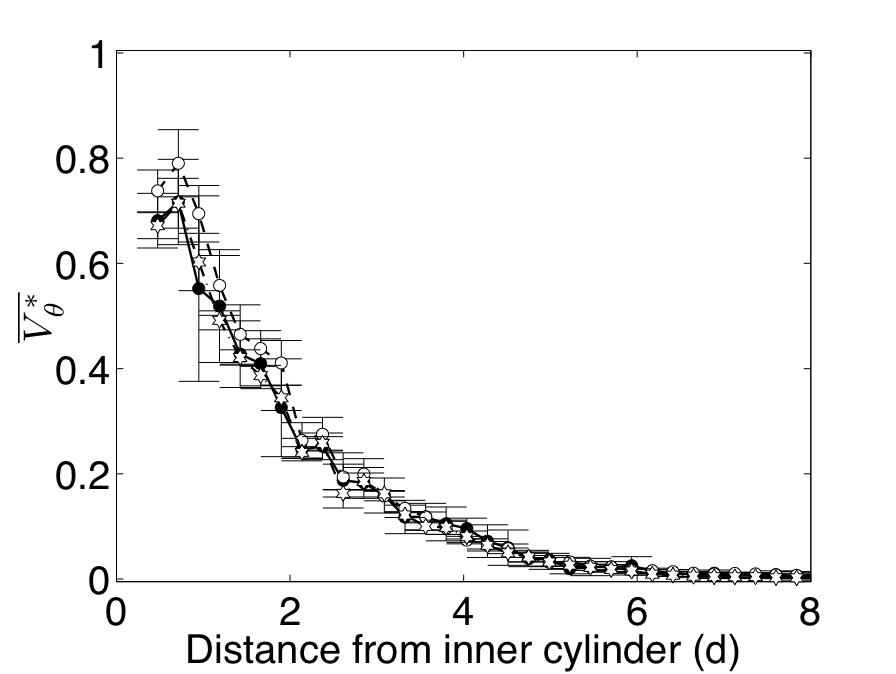}
  \label{fig:Cam2_4mm_MeanAngVelProfs0g}
}
 \caption{Comparison of angular velocity profiles of the 4 mm particles in 1 $g$ and low-gravity.  $\overline{V^*_{\theta}}$ (see \eqn{VstarTheta}) plotted as a function of distance from the inner cylinder for the top surface of the (a) ground-based and (b) microgravity experiments. Mean velocity profiles are shown for the ground based experiments with inner cylinder angular velocities of 0.025 rad s$^{-1}$ (filled triangles) and 0.05 rad s$^{-1}$ (open triangles) and for the microgravity experiments at 0.025 rad s$^{-1}$ (filled circles), 0.05 rad s$^{-1}$ (open circles) and 0.1 rad s$^{-1}$ (open stars). The error bars represent the standard deviation of $\overline{V^*_{\theta}}$ for each group of experiments. The velocity profiles shown only extend up to 8 particle diameters from the shearing surface as further away there is very little particle motion. }
   \label{fig:Cam2_4mm_MeanAngVelProfs}
\end{figure*}

On the top surface of our ground-based experiments (\fig{Cam2_4mm_MeanAngVelProfs1g}) we observe that the mean particle angular velocity ($\overline{V^*_{\theta}}$) decays exponentially with distance from the shearing surface, the shear band is approximately 6-7 particle diameters wide, and the mean normalised angular velocity profiles at different inner cylinder angular velocities are identical to within the error bars.  These results are in agreement with those of many previous ground-based experiments of granular shear with similar experimental set-ups: the angular velocity of particles decreases quickly over a few particle diameters away from the shearing wall and the angular velocity profile, normalised by the shear rate, is independent of the shear rate \citep[\eg][]{behringer99,mueth00, losert00, bocquet01}.  Our microgravity experiments (\fig{Cam2_4mm_MeanAngVelProfs0g}) display the same trends, therefore, we show for the first time, that the same is also true for granular shear in microgravity. 

\begin{center}
  [FIGURE \ref{fig:Cam2_4mm_MeanAngVelProfs} GOES HERE]
\end{center}

\begin{figure*} 
\centering
\subfigure[]{
 \includegraphics[width=0.45\columnwidth]{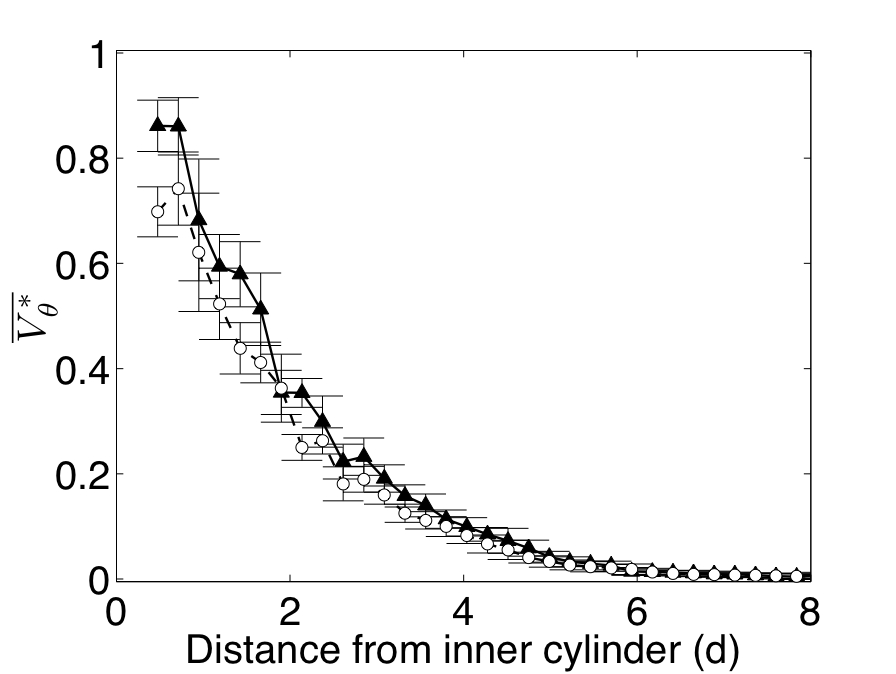}
 \label{fig:Cam2MeanAngVelProfsALL}
}
\subfigure[]{
 \includegraphics[width=0.45\columnwidth]{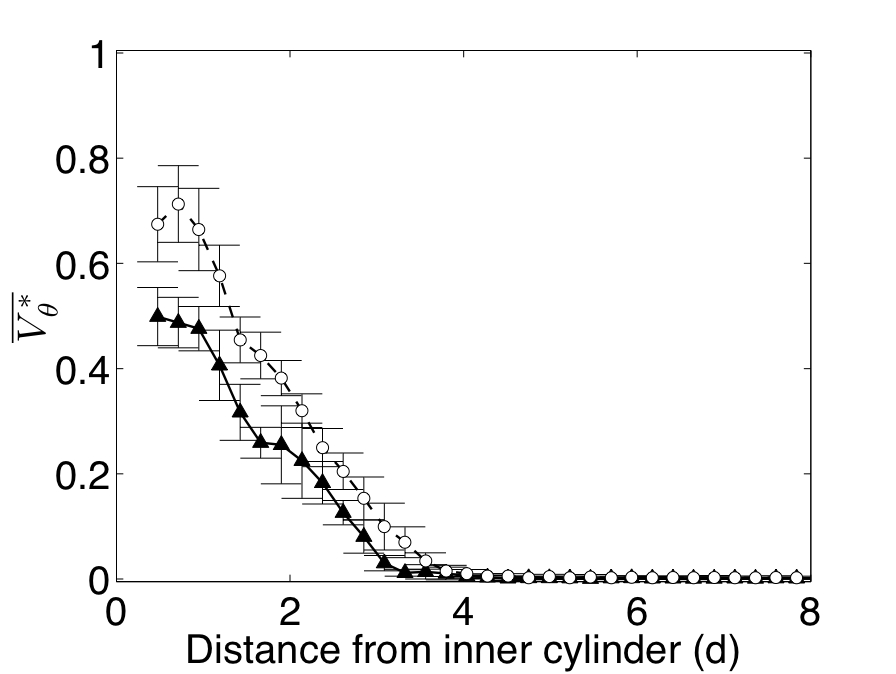}
 \label{fig:Cam1MeanAngVelProfsALL}
 }
 \caption{Comparison of angular velocity profiles of the 4 mm particles (a) the top surface and (b) the bottom surface.  $\overline{V^*_{\theta}}$ is plotted as a function of distance from the inner cylinder. There are two types of experiment: ground-based (black diamonds) and microgravity (open circles). For both types, all experiments at all rotational velocities have been combined to produce the mean $\overline{V^*_{\theta}}$. The error bars represent the standard deviation of $\overline{V^*_{\theta}}$ for each group of experiments.}
  \label{fig:4mmALLAngProfs}
\end{figure*}

Comparing the shape of the top surface angular velocity profiles obtained on the ground and in microgravity shows that the mean normalised angular velocity profiles on the top surface are identical in both microgravity and on the ground, except for the small difference in magnitude near the inner cylinder (\fig{Cam2MeanAngVelProfsALL}). This difference can be explained by considering the shear cell set-up. The outer wall of the inner cylinder is coated with beads to make it rough. These beads are glued onto the wall of the inner cylinder but, due to the presence of the pressure plate, cannot extend above the height of the top layer of beads in the shear cell. Therefore, as the granular material dilates in the microgravity environment, the particles on the top surface nearest the inner cylinder will be in contact with the smooth surface of the inner cylinder rather than the rough surface of the glued-on beads. This will result in a reduced particle velocity very close to the inner cylinder in microgravity compared to the velocity on the ground. 

\begin{figure*} 
\centering
\subfigure[]{
\includegraphics[width=0.45\columnwidth]{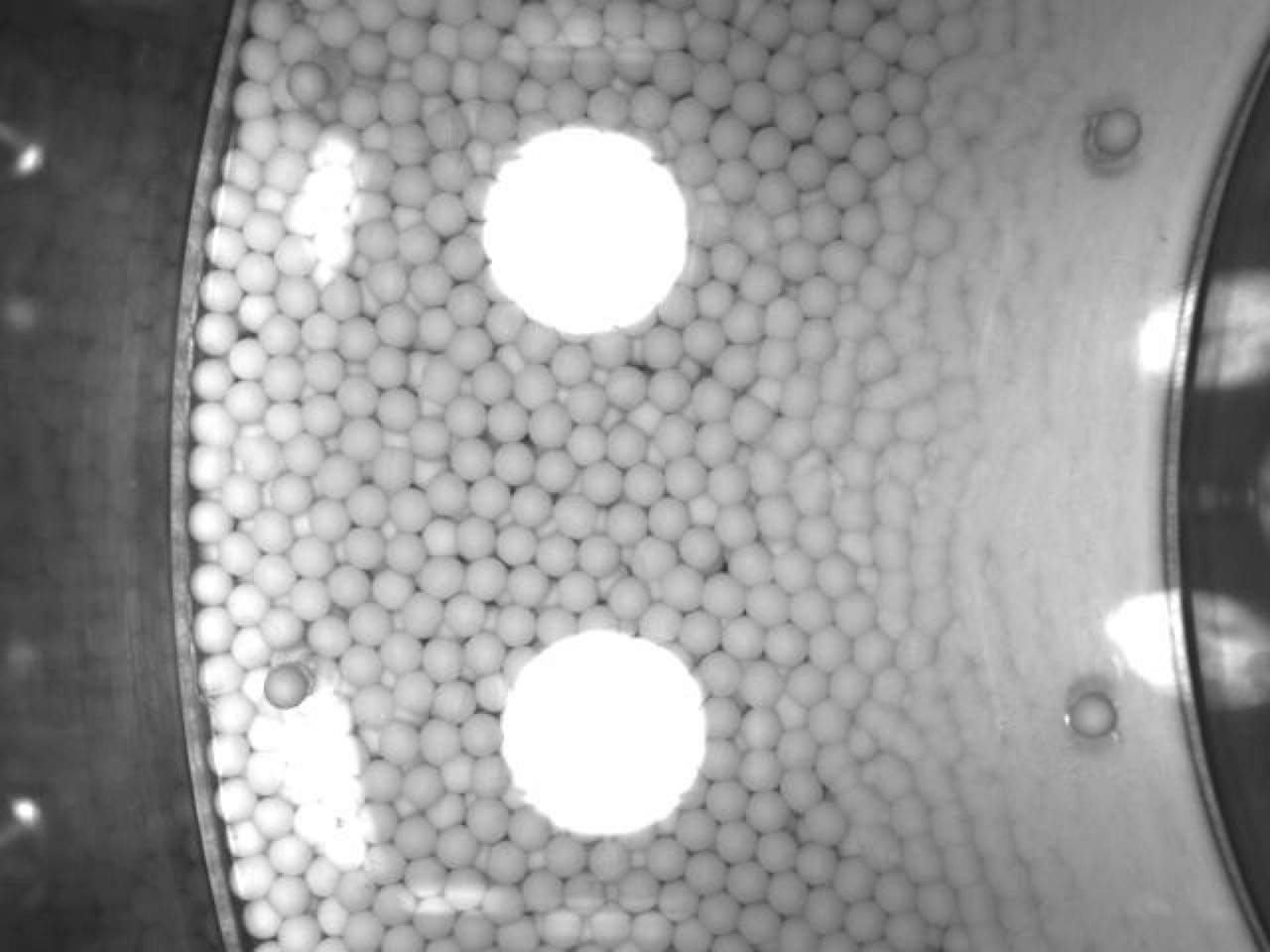}
  \label{fig:StackedAbove}
}
\subfigure[]{
\includegraphics[width=0.45\columnwidth]{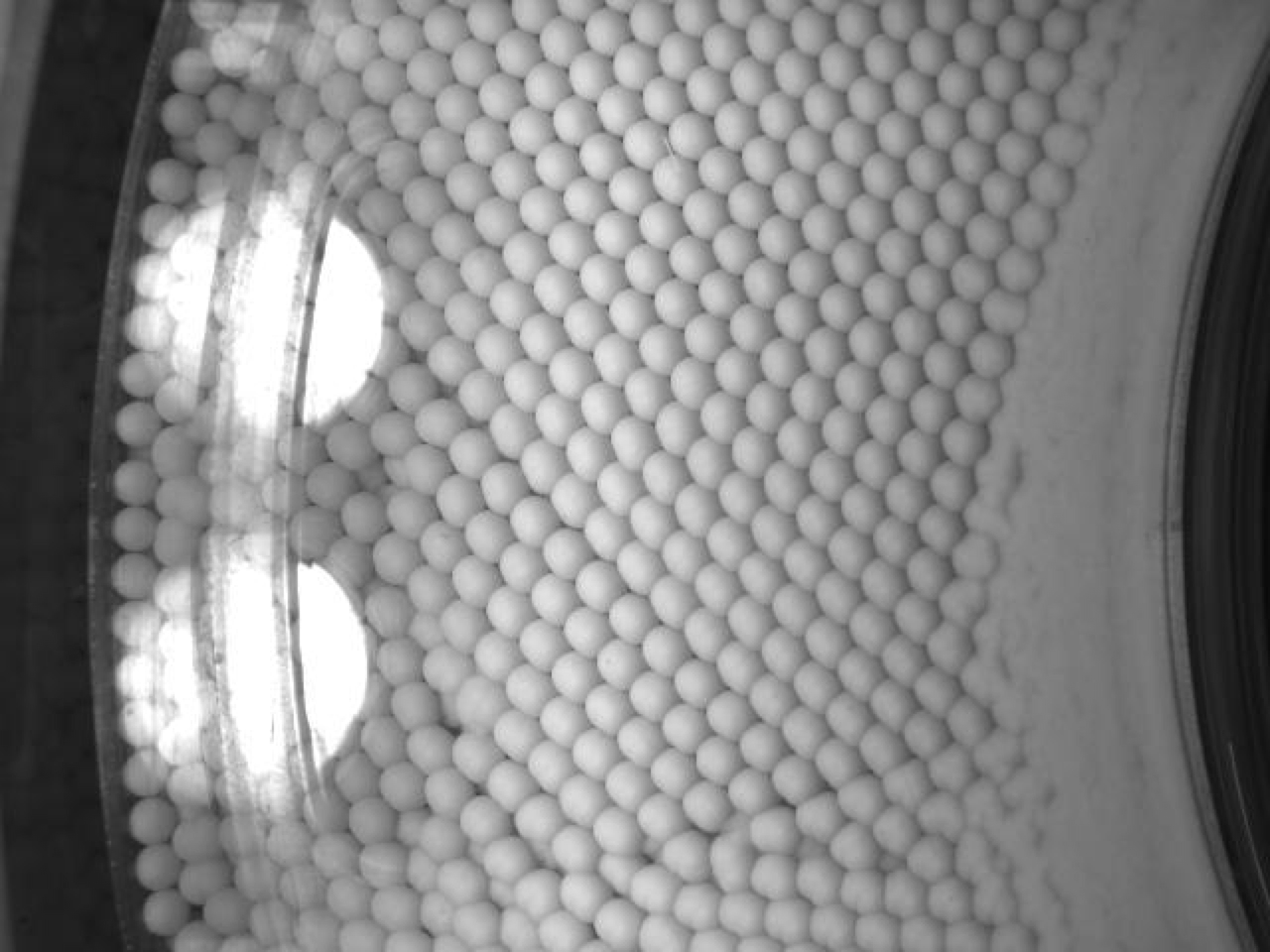}
  \label{fig:StackedBelow}
}
 \caption{Superposition of experimental images. Stacked image made from $>$4000 frames taken of (a) the top surface, and (b) the bottom surface of the granular material. Part of the inner cylinder wall is seen on the right and part of the outer cylinder wall is seen on the left. The two large bright white spots are the reflections of the lamps on the pressure plate. Further reflections can also be seen on the top surface close to the outer cylinder and close to the inner cylinder. In (a) the four beads that are glued to the top surface of the confining pressure plate can also be seen. On the right of both images, close to the inner cylinder, a shear band can be seen. This shear band is much larger in the disordered granular material of the top surface than in the crystallised granular material on the bottom surface.}
   \label{fig:StackedImages}
\end{figure*}

The same trends occur on the bottom surface of the shear cell on the ground and in microgravity (\fig{Cam1MeanAngVelProfsALL}), however, the shear band is much narrower (3-4 particle diameters wide) and the angular velocity profile is very steep.  It is possible that the difference in the angular velocity profiles results from a difference in the packing fraction between the free top surface and the crystallised (\ie hexagonally packed) bottom surface. Indeed, this same effect was found by \cite{daniels06} who reported that when a granular material is in the disordered state the shear band extends to several particles, but while in the crystallised state, the shear is localised almost entirely to the first layer of particles.  This hypothesis gains strength when stacked images of the top and bottom surfaces are considered in which several thousand images have been super-imposed (\fig{StackedImages}). The shear band near to the moving inner cylinder can be seen on the right of both images but it is much smaller in the crystallised granular material on the bottom surface than on the top surface. As a narrow shear band is observed in microgravity as well as on the ground, this would imply that the crystallisation continues to infuence the particle dynamics in microgravity.

It is also worth noting that the magnitudes of the angular velocity profiles at the bottom surface (\fig{Cam1MeanAngVelProfsALL}) are much lower than at the top surface (\fig{Cam2MeanAngVelProfsALL}).  This may be linked to the crystallisation of the bottom surface but it is perhaps more likely to be because, in 1 $g$, the particle motion on the bottom surface is resticted due to the weight of the particles above (recall there are $\sim$13.5 kg of beads in the shear cell; see \sect{MM_cell}).  The same difference in magnitude of $\overline{V^*_{\theta}}$ on the top and bottom surfaces does not exist for the 3 mm beads because the motion of the top surface is also inhibited, albeit via a different mechanism.

Although not shown here, the 3 mm particles follow all of the same trends mentioned above, however, the magnitudes of the mean particle velocites on the top surface are smaller due to the presence of the pressure plate (see \sect{pp}).

\begin{center}
  [FIGURE \ref{fig:4mmALLAngProfs} GOES HERE]
\end{center}

\begin{center}
  [FIGURE \ref{fig:StackedImages} GOES HERE]
\end{center}

\section{Conclusions}

We have designed and flown a Taylor-Couette shear cell to investigate granular flow due to rotational shear forces in varying gravitational environments. The technical details of the experiment have been presented along with our first experimental results. The experiments occur under weak compression. From our experiments and analysis of steady state granular shear in varying gravitational environments, we have shown that:

\begin{itemize}
\item{The angular velocity of particles decreases quickly over a few particle diameters away from the shearing wall for experiments performed both on the ground and in microgravity.}
\item{The normalised angular velocity profiles of constant shear rate experiments performed both on the ground and in microgravity are independent of shear rate.}
\item{The normalised angular velocity profiles of constant shear rate experiments performed both on the ground and in microgravity are almost identical (except for a few differences caused by the experiment set-up).}
\item{A higher packing density of particles (such as at the bottom surface) reduces both the width of the shear band and the maximum angular velocity of the particles in experiments performed both on the ground and in microgravity.}
\end{itemize}

Taking all of the above into account we conclude that the effect of constant shearing on a granular material in a direction perpendicular to the effective acceleration does not seem to be strongly influenced by gravity. This means that shear bands can form in the presence of a weak gravity field just as on Earth. 

These are the first experimental results reporting the response of granular material to shear forces in a microgravity environment. Therefore, although most asteroids have a gravitational acceleration at their surface that is even lower than the levels of gravity obtained in this experiment, these results may be of interest for interpreting images of asteroid surfaces and in the design of future asteroid space missions.

However, we note that shear depends on many factors, for example, wall friction, the effective pressure forcing particles against the side boundaries, the velocity difference between the particles and the walls, the grain characteristics \citep{knight97, mair02, luding08}. There are many more experiments, or indeed numerical simulations, which could be performed to investigate the influence of these factors in varying gravitational environments. 

Further parabolic flights experiments have been performed with our Taylor Couette shear cell investigating granular convection and shear reversal in varying gravitational environments. These experiments are currently being analysed and the results will be presented in future papers.

\section*{Acknowledgments}
Thanks to The Open University, Thales Alenia Space, the UK Science and Technology Facilities Council, the Royal Astronomical Society and the French National Programme for Planetology for providing financial support. Thank you also to ESA ÔFly your ThesisÕ for giving us the opportunity to be part of the 51st ESA microgravity research campaign, and for the financial support. Finally, thanks to the workshop of the Planetary and Space Sciences Research Institite at The Open University for constructing our experimental hardware. This study benefited from discussions with the International Team (\#202) sponsored by the International Space Science Institute (ISSI), Bern, Switzerland.

\bibliographystyle{spbasic}      
\bibliography{Murdochetal2012_GranularMatter.bib} 

\newpage
\section*{FIGURES}

\newpage

\end{document}